# Polarization vector conjugation in a photonic integrated circuit

Anna J. Miller[1], Carson G. Valdez[1], Anne R. Kroo[1], Marko Šimić[1,2], Marek Vlk[1,3], Charles Roques-Carmes[1,4], David A.B. Miller[1], and Olav Solgaard[1]

[1] *Edward L. Ginzton Laboratory, Stanford University, Stanford, CA 94305, USA*
[2] *University of Graz and Christian Doppler for Structured Matter Based Sensing, Graz 8010, Austria*
[3] *Department of Physics and Technology, UiT The Arctic University of Norway, NO-9037 Tromsø*
[4] *Institute of Science and Technology Austria (ISTA), Klosterneuburg 3400, Austria*

We show how to perform polarization conjugation of light in a simple apparatus without any nonlinear optical materials, based on self-configuring Mach-Zehnder interferometers fed by a polarization splitter. This allows us to completely and automatically remove the polarization variations on a single beam or mode propagating through an optical system with arbitrary birefringence, such as a conventional optical fiber, and compensate polarization variations experienced in the forward direction by propagating the conjugated beam back through the system and recreating the original polarization at the source. We demonstrate this polarization conjugation in an optical fiber setup in different orientations with a measured polarization extinction ratio of -30 dB for the returned beam.

Conventional optical phase conjugation (OPC), sometimes referred to as wavefront reversal, is a process that allows for the removal of spatial aberrations to a beam when propagating back through some system of distorting optics, recreating the original beam at the source [1]. This technique has been used in a wide range of applications, including the correction of aberrations in imaging systems, maintenance of signal integrity in communication, beam quality improvements in high-power lasers, and focusing in highly scattering media such as biological tissue. Most approaches require nonlinear optical materials and external pump lasers to generate the phase conjugate, making high efficiency phase conjugation a significant challenge. Additional complications arise when performing full vector phase conjugation [2-9], which, in addition to reversing the wavefront on the backward propagating beam, also conjugates the polarization state.

The advantage of polarization conjugation can be seen when considering a reciprocal, optical system that alters the polarization state of light as it propagates through the system. This is a behavior observed in many optical systems and components, with a good example being single-mode fiber. If polarization conjugation was performed at the far end of the fiber and the conjugated beam transmitted back through the fiber, then the original polarization state would be automatically restored at the source regardless of the actual polarization transformation imparted by the system. Creating a true full vector phase conjugating mirror capable of such polarization conjugation through nonlinear optics, however, has required specific nonlinear materials and interactions or polarized pump waves, while still suffering from the efficiency challenges of conventional OPC systems. Consequently, few convenient implementations of polarization conjugators have been realized.

More recently, digital optical phase conjugation (DOPC) systems using a digitally controlled loop between CCD cameras or sensors and spatial light modulators (SLM), deformable mirrors, or other similar components have been developed [10-14]. These techniques bypass the nonlinear material requirement and decouple the powers of the conjugated beam and input signal, but at the cost of higher latency and complexity, often requiring extensive calculations and precisely calibrated systems to run reliably. Further, DOPC tends to be polarization-dependent, and full polarization phase conjugation demands separate recording steps for the orthogonal polarization components. In contrast, we propose the implementation of simple, automatic polarization conjugation using a polarization-sensitive, self-optimizing silicon photonic integrated circuit (PIC).

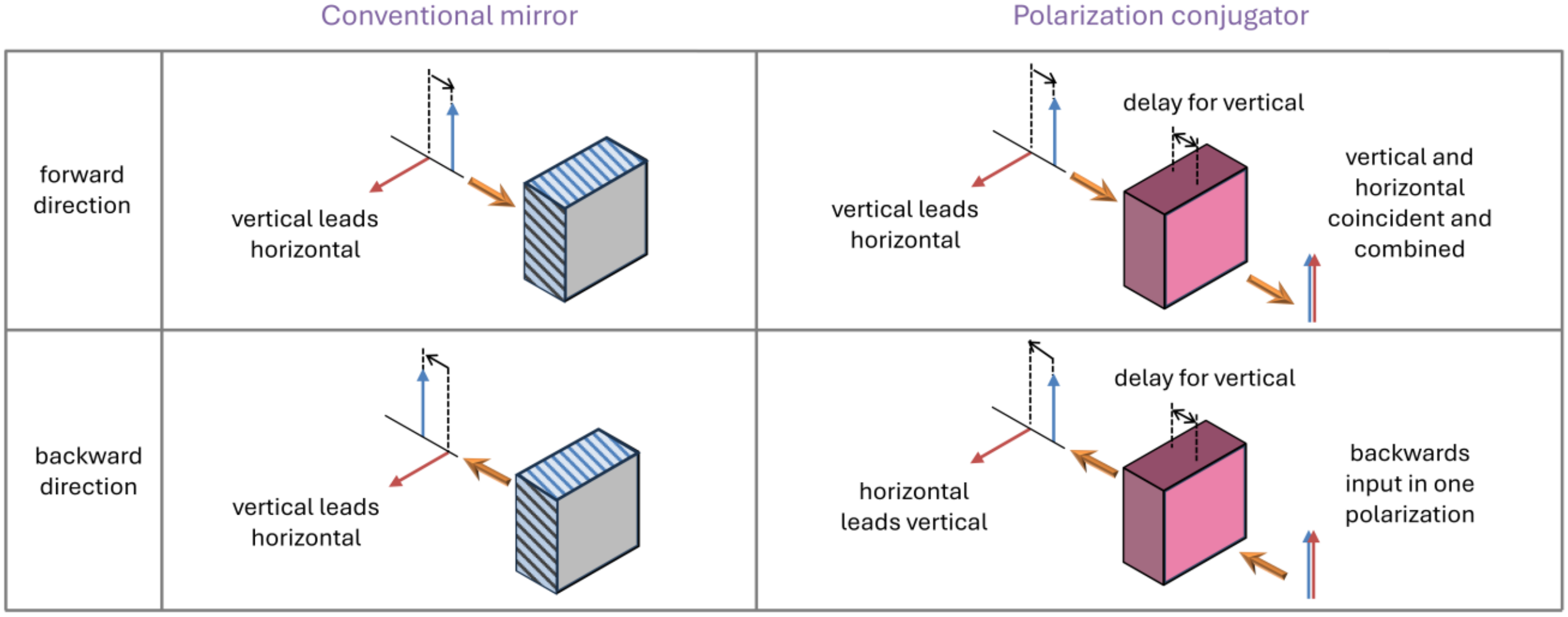

**Fig. 1. Physics of polarization conjugation.** Comparison of the behavior of a conventional mirror and a polarization conjugator (PC). The conventional mirror retains the relative delay between the vertical and horizontal field components on reflection. The PC delays the vertical component so that it is coincident with the horizontal at the output (and both components are rotated to the same polarization). In the backward direction, the field components are split, and the same delay is applied to the vertical component; so, now the horizontal component leads the vertical, resulting in polarization conjugation.

Conceptually, polarization conjugation can be understood from the illustration in Fig. 1. Assuming a general, elliptically polarized incident beam, the "vertical" component will lag or lead the "horizontal" component by some phase. When reflected off an ideal, conventional mirror, the phase relationship is maintained such that the "vertical" component still leads the "horizontal" component, but the propagation direction is flipped. Because of the change in propagation direction, the polarization experiences a change of handedness (e.g., right circular becomes left circular); whereas the horizontal component points to the right in the incident beam, when looking along the propagation direction, in the conventionally reflected beam, it points to the left when looking in the "backwards" propagation direction. Conversely, a phase conjugating mirror will retain the handedness of the input polarization state, even as the propagation direction is reversed, by delaying the "vertical" component relative to the "horizontal" by the same phase amount as it initially led by.

Here, we exploit a polarization splitting grating coupler (PSGC) [15] to separate two orthogonal input polarization components and convert them into two spatial modes of the same polarization in distinct waveguides. This enables the use of a waveguide Mach-Zehnder interferometer (MZI) to both combine those powers into a single output guide by an automatic maximization electronic feedback loop and, in the same operation, to configure the system to emit the polarization conjugate when we shine light back into that output, leveraging the mathematical conjugating property of such linear, lossless networks that can also be utilized in the spatial domain [16, 17]. With this architecture, we demonstrate a compact, low-loss method of polarization conjugation without requiring any nonlinear interactions. The practical availability of polarization state conjugation could counteract many problems commonly associated with polarization in optical systems; isolating any undesired, system-based polarization effects from a signal in this way has considerable appeal in many fiber and free-space applications, such as fiber-optic gyroscopes [18], fiber-based, free-space optical, and quantum communication and encryption [19-22], and enhanced biomedical imaging [23] by contributing to polarization stability and enabling polarization-resolved functionalities in sensing or multiplexing.

## Polarization Conjugation - Concept

The device concept is depicted in Fig. 2 (a). Light normally incident on a polarization splitting grating coupler is decomposed into two orthogonal polarization components, both of which are converted to quasi-TE modes and coupled to separate waveguides [15, 24, 25]. Presuming the light had been completely polarized and coherent, the MZIs can be aligned such that the light emerges entirely from the top output using a self-configuring algorithm [26, 27]. In this process, $\phi$ and $\theta$ are iteratively adjusted to minimize (maximize) the power emerging from the top (bottom) output. Now, light coupled into the top output will propagate backwards through the specified mesh and emerge from the PSGC as the polarization conjugate of the original beam. In this work, the actual device, depicted in Fig. 2 (b,c), uses a symmetric four-port PSGC operating at normal incidence to couple each orthogonal polarization component into separate pairs of waveguides oriented along opposite sides of the PSGC with zero relative phase [15]. The four waveguides then serve as spatial inputs to a two-stage, single-layer binary tree photonic mesh, a minimal architecture that is compatible with the self-configuration algorithms used.

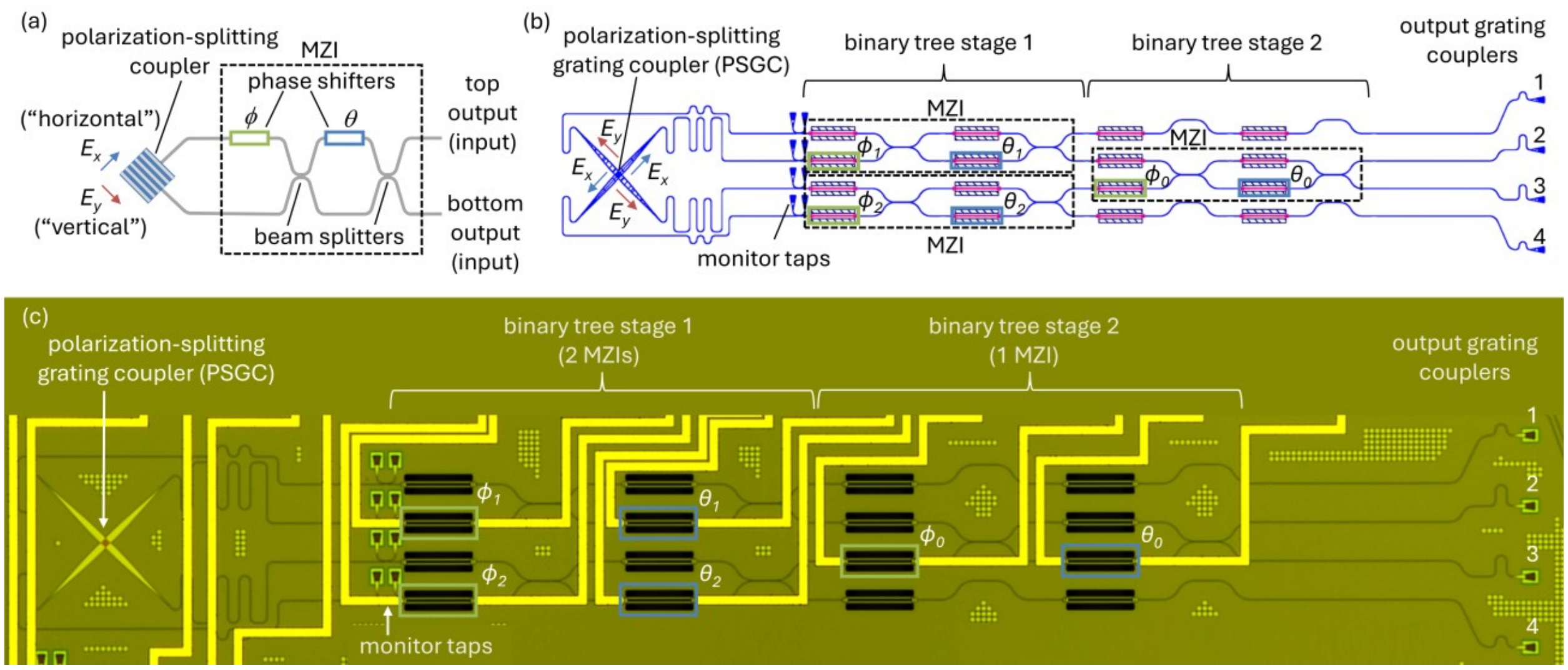

**Fig. 2. Device schematic for on-chip polarization conjugation.** (a) Simplified device concept with an MZI combining the orthogonal polarization components from two outputs of a PSGC (now both TE polarizations in the waveguides). (b) Schematic of the device used in our experiment with a four-port PSGC and two-stage binary tree mesh of MZIs. (c) Annotated microscope image of the fabricated chip.

We can formally describe the process of polarization conjugation in our device mathematically. For simplicity, consider the device as illustrated in Fig. 2 (a). The input to the overall system can be described as a normalized electric field vector, $|\psi_{\mathrm{fwd}}\rangle$, in a single-spatial-mode written in terms of its horizontal ($E_{\mathrm{x}}$) and vertical ($E_{\mathrm{y}}$) amplitude components where only the relative delay between the components is considered

$$|\psi_{\mathrm{fwd}}\rangle \equiv \begin{bmatrix} E_{\mathrm{x}} \\ E_{\mathrm{y}} \end{bmatrix} \equiv \begin{bmatrix} E_{\mathrm{x0}} \\ E_{\mathrm{y0}}\, e^{i\phi_0} \end{bmatrix} \tag{1}$$

Since common phase factors can be neglected, $E_{\mathrm{x0}}$ and $E_{\mathrm{y0}}$ are chosen to be real, positive numbers. By normalization,

$$|E_{\mathrm{x}}|^2 + \left|E_{\mathrm{y}}\right|^2 = 1 \tag{2}$$

the input vector now resembles the Jones vector of the incident field. The behavior of the optical system and conjugator is mapped to its mathematical description in Fig. 3. As the input propagates through some arbitrary, birefringent optical system, represented by the complex 2x2 Jones matrix $U_{\mathrm{OS}}$, the polarization state described by $|\psi_{\mathrm{fwd}}\rangle$ will be altered. When the light is coupled to the PIC, the PSGC converts $|\psi_{\mathrm{fwd}}\rangle$ from orthogonal components in a single spatial mode polarization basis to two distinct spatial waveguide modes in a single polarization state; this is the input into the photonic mesh. Referring to Fig. 2 (a), this translates to a 2x1 vector containing the complex amplitudes in each waveguide.

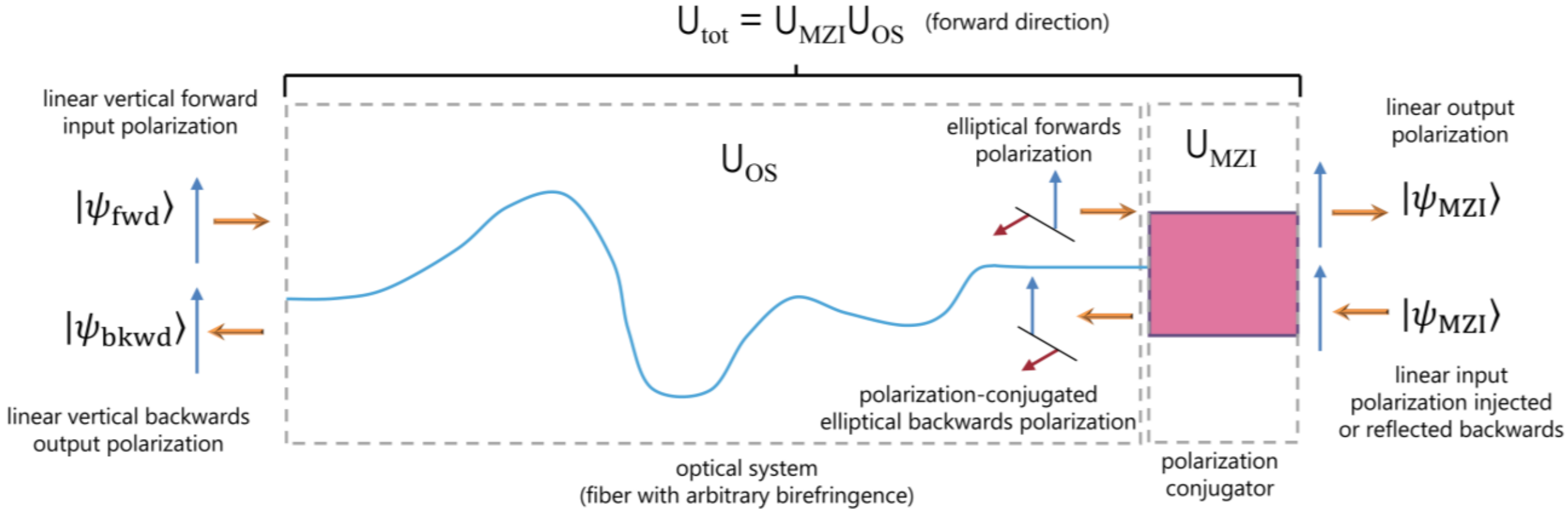

**Fig. 3. illustration of a birefringent fiber example**. The total system is decomposed into two separate unitary matrices: $U_{OS}$, corresponding to the perturbing optical system (in this case, a fiber), and $U_{MZI}$, corresponding to the transformation applied by the MZI between the input and output waveguides. Physically, a PSGC converts the input from a polarization basis to a spatial basis between the optical system and MZI.

At this point, self-configuration is applied to the MZI to interferometrically concentrate all power in a single port (top output), with the phase shifter settings determined by the particular polarization state. Assuming the system is lossless, the cumulative effect of the initial optical system and our device can be described using the normalization in Eq. 2 as

$$U_{\mathrm{MZI}} U_{\mathrm{OS}} |\psi_{\mathrm{fwd}}\rangle \equiv |\psi_{\mathrm{MZI}}\rangle = \begin{bmatrix} 1 \\ 0 \end{bmatrix} \tag{3}$$

where the upper element of the output vector, $|\psi_{\mathrm{MZI}}\rangle$, is the amplitude emerging from the top port of the MZI.

The direction of propagation is then reversed by coupling $|\psi_{\mathrm{MZI}}\rangle$ as defined in Eq. 3 to the output ports. Since we are treating the overall system, $U_{MZI}U_{OS}$, as unitary, this reverse operation can be described as the transpose because of electromagnetic reciprocity [22]. As a result, the backward-propagating amplitude, $|\psi_{\mathrm{bkwd}}\rangle$, emerging at the original input of the system is represented as

$$|\psi_{\mathrm{bkwd}}\rangle = (\mathrm{U}_{\mathrm{MZI}} U_{\mathrm{OS}})^T |\psi_{\mathrm{MZI}}\rangle \tag{4}$$

Taking the complex conjugate of all terms in Eq. 4 and using the fact that $|\psi_{\mathrm{MZI}}\rangle$ is real,

$$|\psi_{\mathrm{bkwd}}\rangle^* = (U_{\mathrm{MZI}} U_{\mathrm{OS}})^\dagger |\psi_{\mathrm{MZI}}\rangle \tag{5}$$

By substituting in Eq. 3, the backward-propagating amplitude is related to the original input as

$$|\psi_{\mathrm{bkwd}}\rangle^* = (U_{\mathrm{MZI}} U_{\mathrm{OS}})^\dagger (U_{\mathrm{MZI}} U_{\mathrm{OS}}) |\psi_{\mathrm{fwd}}\rangle \tag{6}$$

Noting that generally for matrix products that $(U_{\mathrm{MZI}} U_{\mathrm{OS}})^\dagger = {U_{\mathrm{OS}}}^\dagger {U_{\mathrm{MZI}}}^\dagger$ and that for a unitary matrix, $U$, $UU^\dagger$ is the identity matrix, then after also taking the complex conjugate once more,

$$|\psi_{\mathrm{bkwd}}\rangle = |\psi_{\mathrm{fwd}}\rangle^* \equiv \begin{bmatrix} E_{\mathrm{x}} \\ E_{\mathrm{y}} \end{bmatrix}^* \tag{7}$$

Therefore, the beam propagating backwards from the original input is the polarization conjugate of the original polarization state. Since no assumptions were made about the optical system, $U_{OS}$, beyond being lossless and reciprocal, this principle applies to any optical system that can be regarded as unitary, making the preservation of phase conjugation a general feature of unitary, reciprocal systems [22]. In this case, we are taking advantage of the fact that the phase conjugate of the field in a single-mode waveguide in a particular polarization is identical to that of the mode going backwards in the same waveguide, so the generation of the phase conjugate is trivial for a single mode in a single polarization.

Although our implementation features a four-input PIC, the device is functionally the same as the two-input concept. The four-port PSGC converts the orthogonal polarization components to the same polarization state in the waveguides, transforming the input from the polarization basis to a four-element spatial basis instead of two elements. The two-element output of our mesh as described in Eq. 3 can be adapted to our four-input/four-output architecture depicted in Fig. 2 (b) by padding the top and bottom of $|\psi_{MZI}\rangle$ with zeros.

In this description we have assumed a lossless system and neglected global phase factors. In a real system, loss is unavoidable, but if the loss is the same for all polarizations, the system can still be described by a unitary matrix multiplied by an overall loss factor. Similarly, adding a common phase factor for the overall propagation through the system will not alter the operations discussed above. Both cases will introduce an overall complex factor to the conjugated polarization but leave the relative amplitudes and phases of the polarization states unaffected, so the polarization conjugating property described under these assumptions is preserved.

**Polarization Conjugation - Experiment**

In the experimental verification, we used standard single-mode fiber (SMF) and a manual polarization controller as an arbitrary optical system with some polarization-scrambling effect as described by the Jones matrix, $U_{OS}$. The polarization conjugating PIC was set up as shown in Fig. 4 (a). A fiber-based polarization beam splitter (PBS) was used to define our polarization basis in terms of two orthogonal orientations as separated by the birefringent prism. To enforce a known polarization state on the input, $|\psi_{fwd}\rangle$, 1550 nm light from a fiber-coupled laser in polarization-maintaining (PM) fiber was connected to one of the linearly polarized arms of the PBS, such that

$$|\psi_{\mathrm{fwd}}\rangle \equiv \begin{bmatrix} E_{\parallel} \\ E_{\perp} \end{bmatrix} = \begin{bmatrix} E_0 \\ 0 \end{bmatrix} \tag{8}$$

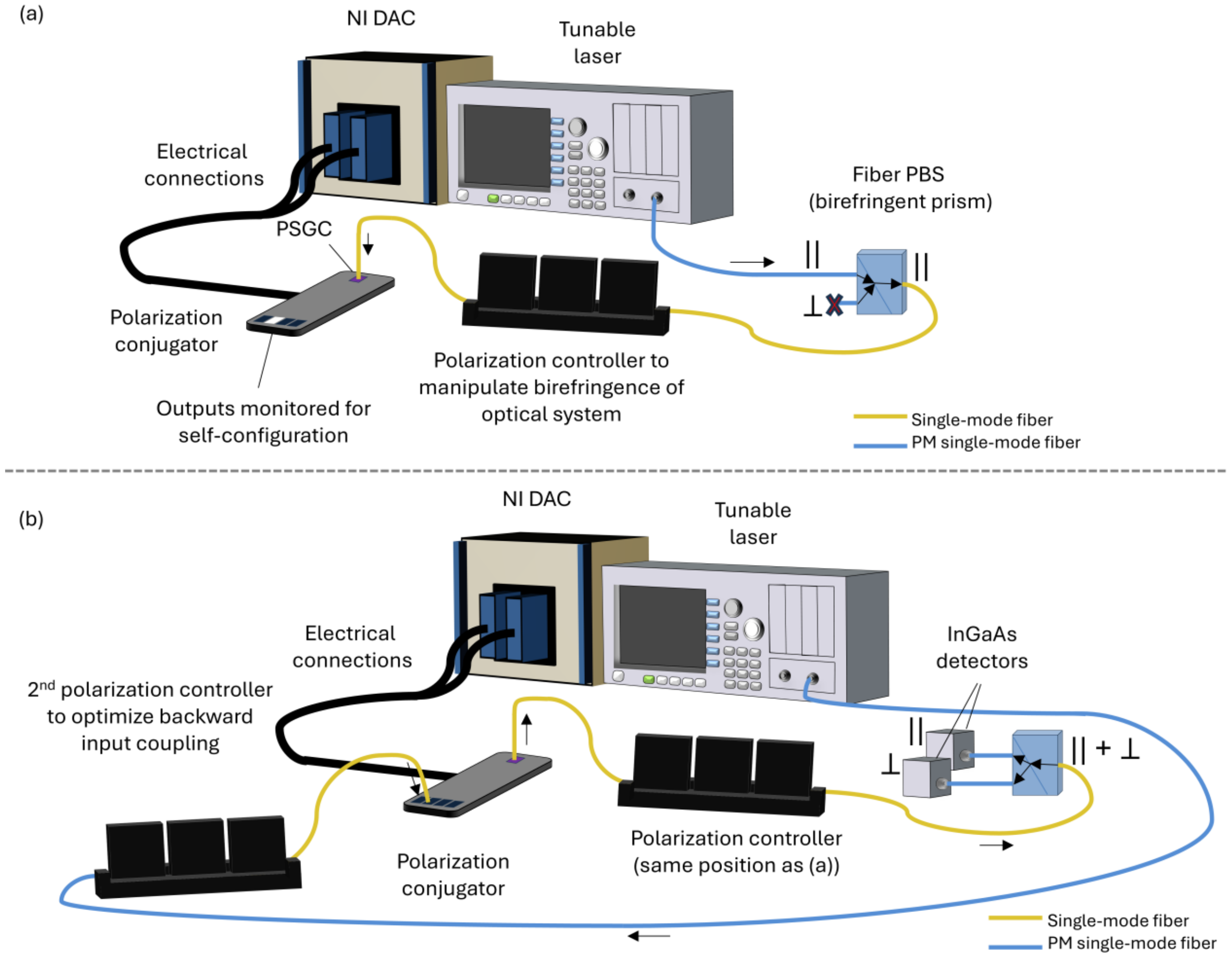

**Fig. 4. Diagram of the experimental setup.** (a) The polarization conjugator receives an input through the PSGC through an arbitrary, birefringent system and self-configures to maximize the power in a single port. (b) The input is switched to the output side to measure the polarization extinction ratio of the generated conjugate beam emitted from the PSGC when operating in the reverse direction.

After passing through the prism, the input was coupled to an SMF and propagated through our birefringent optical system before being coupled to the PIC by the PSGC. To maximize the light from output 2 as defined in Fig. 2 (b), we configured the mesh using a dithering approach and gradient descent [28, 29]. This was accomplished by simultaneously modulating each phase shifter voltage at a unique frequency within an octave and monitoring the power at output 2. The gradient was calculated from this and fed into a gradient-descent optimizer [30]. Typically, convergence was seen within 10-15 iterations as shown in Supplementary Fig. 1.

We then reversed the direction of propagation by connecting the laser to a separate fiber to couple light into output 2 as shown in Fig. 4 (b). Since the birefringent optical system between the PBS and PIC was unchanged, perfect polarization conjugation would result in an identical polarization state as the original emerging backwards from the PBS, which in this case is

$$|\psi_{\mathrm{bkwd}}\rangle = \begin{bmatrix} E_0^* \\ 0 \end{bmatrix} \tag{9}$$

To quantify this, we measured the power from both input arms of the PBS in reverse operation with reference to the original polarization and calculated the polarization extinction ratio (PER) as the ratio of intensities between the linearly polarized component aligned with the original input, $I_{||}$, and the orthogonal linearly polarized component, $I_{\perp}$.

$$PER = 10\log_{10}\left(\frac{I_{\parallel}}{I_{\perp}}\right) \tag{10}$$

This process was repeated for a collection of arbitrary paddle orientations on the polarization controller, effectively changing $U_{OS}$. During the self-configuration process, we measured the polarization state of the input to the PSGC [15] and essentially determined $U_{OS}$, which can be extracted from the phase shifter settings. The measured polarization states and corresponding extinction ratios are plotted on the Poincaré sphere in Fig. 5 (a). Since the input state was a known linear polarization, the amount of polarization mixing in the fiber can be quantified by taking the inner product of the normalized input Jones vector, $\psi_{||}$, and the measured normalized Jones vector, $\psi_{\mathrm{conj}}$. In Fig. 5 (b), the extinction ratios are plotted with respect to the inner product. For all fiber orientations, the calculated PER remained between -28.0 dB and -32.7 dB with an average ratio of -30.4 dB, showing that the device's ability to compensate for a birefringent optical system is substantially independent of the specific transformation performed by the optical system.

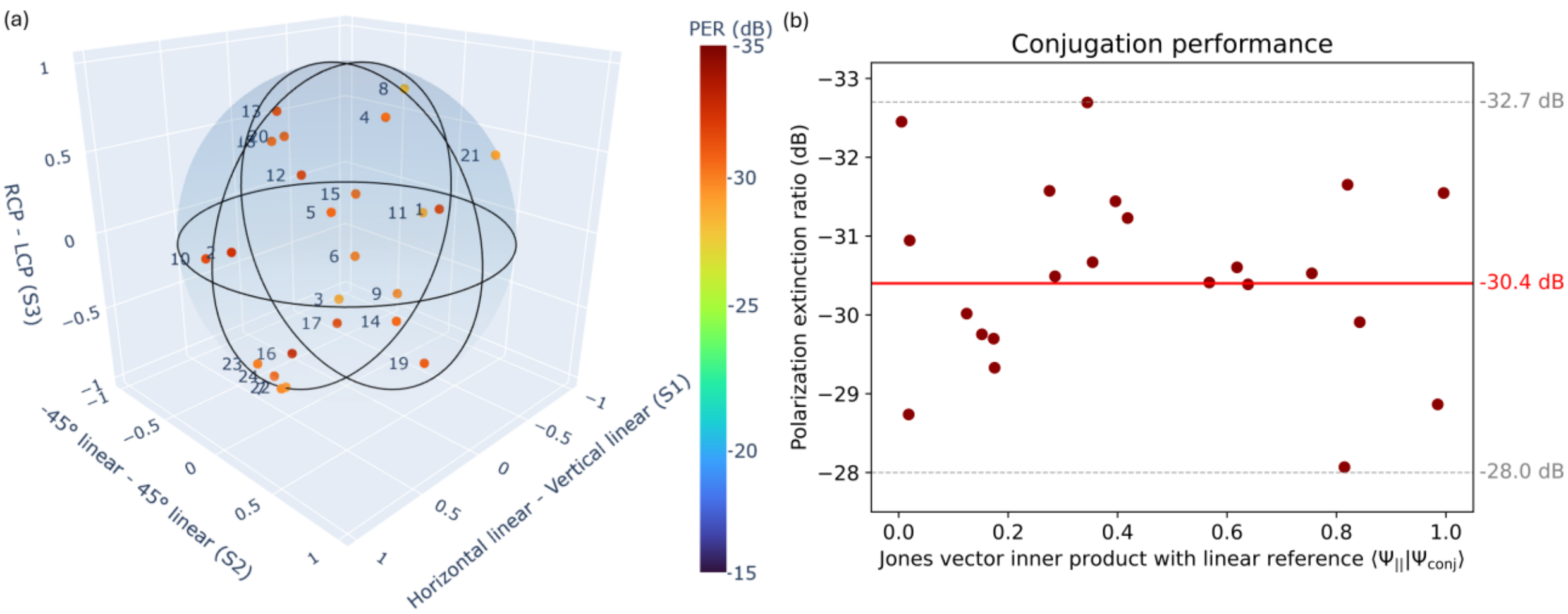

**Fig. 5. PER measured for a set of 24 arbitrary fiber configurations.** (a) PER plotted on the Poincaré sphere representing the polarization state of the beam as received at the PSGC after propagating through the fiber optical system. (b) PER plotted individually for each fiber configuration. The x-axis gives a measure of polarization mixing present in the fiber by reporting the inner product of the original linear Jones vector ($\psi_{||}$) and the Jones vector measured after exiting the fiber ($\psi_{\mathrm{conj}}$).

The performance of our polarization conjugator was also evaluated across the wavelength range 1500 nm – 1580 nm using the same experimental setup and procedure. This experiment showed optimal performance between 1530 nm and 1550 nm which corresponds to the expected behavior of the mesh and the PSGC [15]. This data is reported in Supplementary Fig. 2.

**Discussion**

We have presented a reconfigurable system comprised of only linear components that can generate conjugated polarization states. The fidelity (F) of the polarization conjugate state, which, in the case of full polarization, is the inner product of a generated state and the target state, can be inferred from $|\psi_{\text{bkwd}}\rangle$. If the returned state at the input is identical to the input, the conjugate state must have been perfectly generated. Perfect conjugation would result in $I_{\perp}= 0$; based on our measurements, we calculate F > 0.99 for each fiber configuration, which compares well with previously reported performance metrics for the conjugation of isotropic polarization states [5, 14].

The ability of our system to compensate for polarization variations along an optical path is primarily limited by two factors. First is the assumption that no polarization dependent loss (PDL) exists in our device. Since the orthogonal polarization components are converted to a single polarization state in the waveguides of the binary tree, we expect that the mesh operates equivalently on each element of the input vector. Therefore, we would attribute any PDL present in the system to the coupling process of light to the PSGC. In this experiment, the PSGC exhibited an inherent PDL that peaks at 1.2 dB and is more prominent at the edges of the operating band [15]. This difference is likely due to slight fabrication errors in the structure that alter the device symmetry. Another source of PDL lies in the alignment of the incident beam to the PSGC. Any lateral or angular offset from the centered, normally incident case can cause deviations in the relative coupling efficiencies of the polarization components to their respective ports, which changes the ratio of the orthogonal components seen by the mesh. It is reasonable to expect that, with knowledge of the PDL of a given PIC, appropriate steps can be taken to further improve performance such as incorporating add/drop ports in the circuit to mitigate the loss difference.

Second, the ability of the mesh to route all light to a single port relies on ideal 50:50 beam splitters in each MZI. Our proof-of-concept device does not necessarily satisfy this condition in its current iteration, but the same architecture can be modified to achieve perfect splitting [31], which would improve the achievable extinction ratio of our conjugator. We estimate that the actual splitting ratio of the MZIs in this PIC is closer to 49.6 : 50.4 and expect that modifying the MZIs could improve this to a nearly perfect splitting ratio of 49.96 : 50.04 [32].

Experimentally, we have demonstrated polarization conjugation through the compensation of birefringence in stationary optical fiber, but it is also possible to use the same device to track polarization aberrations in a time-varying system and compensate them. The bidirectionality of the mesh can be used to self-configure on some reference input and generate a conjugated beam in the opposite direction at the same time. The environmental birefringence effects typically observed in optical systems tend to drift on a millisecond to second time scale. The PIC in our experiment was not operated at high speeds; each iteration in our self-configuration algorithm took roughly 1.18 seconds, with convergence occurring within 15 iterations. This, however, is not an architectural limitation. With the proper electronics, phase shifter settings in silicon photonic meshes have been shown to converge in tens of milliseconds using self-configuration algorithms, with speeds limited by the response time of the phase shifters [33]. This can be further enhanced by swapping out the thermal phase shifters for faster actuators, such as micro-electromechanical systems (MEMS), carrier-depletion methods, or electro-optic materials. Compared to earlier techniques using photorefractive crystals [10, 11] or adaptive digital approaches [12-14], this represents a significantly faster method of polarization conjugation. Granted, our approach cannot operate as fast as nonlinear processes [5], but we expect it can achieve rates more than sufficient for most practical applications without requiring external sources or nonlinear materials. While nonlinear methods require powerful pump lasers, our approach can generate a reflected conjugate beam using only the input by placing a mirror above the output grating coupler such that the output is reflected back into the grating coupler and propagated through the mesh to the original source. With a high efficiency grating coupler, we can minimize coupling loss for the signal and return a conjugated beam comparable to the original input in strength.

In this work, our primary focus was on demonstrating the polarization conjugating ability of the mesh, but because phase conjugation is a general property of unitary meshes, full vector phase conjugation is a natural extension. Self-configuring silicon photonic meshes with an array of input grating coupler can be used to analyze and generate spatial modes [34, 35] where each grating coupler acts as a pixel of the beam. By using polarization-splitting inputs in a spatial array, the mesh could capture both polarization information and spatial phase relationships between each grating. It is worth noting that we can only conjugate as many spatial modes as the mesh can resolve, which is limited to the number of inputs to the mesh. Operating such a configured mesh in reverse would satisfy both criteria for ideal full vector phase conjugation by conjugating both the polarization vector and wavefront.

In conclusion, we have shown an integrated architecture and approach to polarization conjugation without nonlinear optics. As in DOPC systems, the nonlinearity of the operation is provided electronically, namely by a feedback loop between the phase shifter voltages and measured output powers in our self-configuring mesh, but in a compact and relatively simple setup. Additionally, the silicon photonic platform is readily integrable with on-chip photonic components and PICs, making it a convenient, practical implementation of automatic polarization conjugation.

## Materials and Methods

Fabrication was performed by Advanced Micro Foundry through a multi-project wafer shuttle run on 220 nm silicon on insulator using standard process steps. The normal incidence PSGC, designed to operate at 1550 nm, featured a partial etch depth of 70 nm, a grating period of 556 nm, and a 50% fill factor and exhibited a minimum insertion loss of -4.5 dB and -4.3 dB for the TE and TM polarization states respectively [15]. The other, conventional grating couplers were designed to accept and emit 1550 nm light at a 12˚ angle of incidence with a minimum insertion loss of -6 dB. Light was routed through our circuit with silicon single-mode waveguides (500 nm x 220 nm). Each MZI was comprised of a 50:50 beam splitter/combiner, a $\theta$ phase shifter, a second 50:50 beam splitter/combiner, and a $\phi$ phase shifter. The beam splitters had a 300 nm coupling gap, 40 μm interaction length, and 35 μm radius of curvature for bends to minimize losses, and titanium nitride heaters were used for the phase shifters. Thermal isolation trenches were etched on either side of each phase shifter to reduce thermal crosstalk. Inactive phase shifters and path-length equalization bends were placed throughout the PIC to maintain relative phase factors, mitigate wavelength dependence, and ensure loss matching upon propagation. We designed our PSGC using Tidy3D's Finite-Difference Time Domain (FDTD) solver.

In our experiment, we used standard single-mode fiber. To couple light on and off the chip, a fiber was cleaved and placed on an angled mount on a Thorlabs Nanomax 600 series 6-axis stage. The fiber was connected to an HP81680A tunable laser when acting as an input and connected to an HP81531A power meter to measure the output. The birefringent effects of the fiber were altered using a manual polarization controller (Thorlabs FPC652). To isolate orthogonal polarization components, we used a Thorlabs PBC1550SM-APC fiber-based beam splitter. All of the phase shifters in the PIC were controlled by a National Instruments (NI) digital-to-analog converter (DAC).

## Acknowledgements

This work has been funded by the Air Force Office of Scientific Research (FA9550-21-1-0312, FA9550-23-1-0307), Ames Research Center (80NSSC24M0033), and Stanford Engineering. Part of this work was performed at nano@stanford RRID:SCR_026695.

## References

[1] R. W. Boyd, *Nonlinear Optics*. Academic Press. (1992).

[2] W. R. Tompkin, M. S. Malcuit, R. W. Boyd, and J. E. Sipe. Polarization properties of phase conjugation by degenerate four-wave mixing in a medium of rigidly held dye molecules. *J. Opt. Soc. Am. B, JOSAB*, 6, 757–760 (1989).

[3] R. W. Boyd, K. R. MacDonald, and M. S. Malcuit. Vector Phase Conjugation And Beam Combining By Multiwave Optical Mixing. *Laser Wavefront Control*, SPIE, 69-80 (1989).

[4] R. P. M. Green, S. Camacho-Lopez, and M. J. Damzen. Experimental investigation of vector phase conjugation in $Nd^{3+}$:YAG. *Opt. Lett., OL*, 21, 1214–1216 (1996).

[5] A. G. De Oliveira *et al.* Real-Time Phase Conjugation of Vector Vortex Beams. *ACS Photonics*, 7, 249–255, (2020).

[6] A. G. de Oliveira, N. Rubiano da Silva, R. Medeiros de Araújo, P. H. Souto Ribeiro, and S. P. Walborn. Quantum Optical Description of Phase Conjugation of Vector Vortex Beams in Stimulated Parametric Down-Conversion. *Phys. Rev. Appl.*, 14, 024048 (2020).

[7] I. McMichael, M. Khoshnevisan, and P. Yeh. Polarization-preserving phase conjugator. *Opt. Lett*., 11, 525-527 (1986).

[8] S.-X. Qian, L.-J. Kong, Y. Li, C. Tu, and H.-T. Wang. Recording and reconstruction of vector fields in a Fe-doped $LiNbO_3$ crystal. *Opt. Lett., OL*, 39, 1917–1920 (2014).

[9] S.-X. Qian, Y. Li, L.-J. Kong, C. Tu, and H.-T. Wang. Phase conjugation of vector fields by degenerate four-wave mixing in a Fe-doped $LiNbO_3$," *Opt. Lett., OL* 39, 4907–4910 (2014).

[10] Y. Dai *et al.* Active compensation of extrinsic polarization errors using adaptive optics. *Opt. Express, OE*, 27, 35797–35810 (2019).

[11] lC. He, J. Antonello, and M. J. Booth. Vectorial adaptive optics. *eLight*, 3, 23 (2023).

[12] M. Cui and C. Yang. Implementation of a digital optical phase conjugation system and its application to study the robustness of turbidity suppression by phase conjugation. *Opt. Express*, 18, 3444-3455 (2010).

[13] Y. Shen, Y. Liu, C. Ma, and L.V. Wang. Focusing light through scattering media by full-polarization digital optical phase conjugation. *Opt. Lett*., *OL*, 41, 1130-1133 (2016).

[14] J. Dou, Y. Mai, W. Jiang, K. Wang, L. Zhong, J. Di, *et al.* Vector modulation of fully-polarized phase conjugate light field through scattering media. *Opt. Laser Technol.* 181, 111987 (2025).

[15] Valdez, C.G., Miller, A.J., Kroo, A.R. *et al.* Integrated photonic polarization synthesizer and analyzer. *Light Sci Appl* 15, 309 (2026).

[16] R. Ulrich and M. Johnson. Fiber-ring interferometer: polarization analysis. Opt. Lett. 4, 152-154 (1979).

[17] I. P. Kaminow, T. Li, and A. E. Willner. *Optical Fiber Telecommunications VB: Systems and Networks*. Academic Press (2010).

[18] C. Agnesi *et al.* Simple quantum key distribution with qubit-based synchronization and a self-compensating polarization encoder. *Optica* 7, 284-290 (2020).

[19] N. Gisin, G. Ribordy, W. Tittel, and H. Zbinden. Quantum cryptography. *Rev. Mod. Phys*. 74, 145–195 (2002).

[20] J. Chen et al. Free-Space Communication Turbulence Compensation by Optical Phase Conjugation. *IEEE Photonics Journal*, 12, 1-11 (2020).

[21] M. Pircher, E. Götzinger, B. Baumann, C.K. Hitzenberger. Corneal birefringence compensation for polarization sensitive optical coherence tomography of the human retina. *J Biomed Opt*. 12, 4 (2007).

[22] D. A. B. Miller. Analyzing and generating multimode optical fields using self-configuring networks. *Optica* 7, 794–801, (2020).

[23] S. SeyedinNavadeh *et al.* Determining the optimal communication channels of arbitrary optical systems using integrated photonic processors. *Nat. Photon.* 18(2), 149–155 (2024).

[24] F. Van Laere, W. Bogaerts, P. Dumon, Gü. Roelkens, D. Van Thourhout, and R. Baets. Focusing Polarization Diversity Grating Couplers in Silicon-on-Insulator. *Journal of Lightwave Technology*, 27(5), 612–618 (2009).

[25] X. Feng, W. Zhou, H. Chen, W. Tian, Z. Chen, Y. Ciu, *et al.* High-Efficiency, Broadband, and Polarization-Diversity Multicore Fiber-Chip Interfaces for Optical I/O. *ACS Photonics*, (2026).

[26] D. A. B. Miller. Self-aligning universal beam coupler. *Opt. Express*, 21(5), 6360–6370 (2013).

[27] D. A. B. Miller. Self-configuring universal linear optical component. *Photon. Res*. 1(1), 1–15 (2013).

[28] F. Zanetto, V. Grimaldi, F. Toso, E. Guglielmi, M. Milanizadeh, D. Aguiar, *et al*. Dithering-based real-time control of cascaded silicon photonic devices by means of non-invasive detectors. *IET Optoelectronics*, 15(2), 111–120 (2021).

[29] P.A. Mor, A.R. Kroo, C.G. Valdez, M. Šimić, A. Karnieli, G. Cavicchioli, *et al*. Separating partially coherent light. arXiv preprint arXiv:2603.15517 (2026).

[30] D. P. Kingma. Adam: A method for stochastic optimization. arXiv preprint arXiv:1412.6980 (2014).

[31] D. A. B. Miller. Perfect optics with imperfect components. *Optica* 2, 747–750 (2015).

[32] C.G. Valdez, Z. Sun, A.R. Kroo, D.A.B. Miller, and O. Solgaard. High-contrast nulling in photonic meshes through architectural redundancy. *Opt. Lett*. 50, 3660-3663 (2025).

[33] F. Zanetto, F. Toso, V. Grimaldi, M. Petrini, A. Martinez, M. Milanizadeh, *et al.* Time-Multiplexed Control of Programmable Silicon Photonic Circuits Enabled by Monolithic CMOS Electronics. *Laser Photonics Rev*,17, 2300124 (2023).

[34] V. Sharma, D. Brandmüller, J. Bütow, J. S. Eismann, and P. Banzer. Universal photonic processor for spatial mode decomposition. *Nat Commun*, 16, 7982 (2025).

[35] J. Bütow, V. Sharma, D. Brandmüller, J. S. Eismann, and P. Banzer. Photonic integrated processor for structured light detection and distinction. *Commun Phys* 6, 369 (2023).